\documentclass[%
    reprint,
    superscriptaddress,
    amsmath,amssymb,
    aps,
    pre,
]{revtex4-2}

\usepackage{graphicx}
\usepackage{dcolumn}
\usepackage{bm}
\usepackage{hyperref}
\usepackage{physics}

\newcommand{\extbes}{\tilde{i}}

\begin{document}

    \title{Evaluation of Gaussian integrals for the modeling of two-dimensional
    quantum systems}

    \author{{\O}yvind Sigmundson Sch{\o}yen}
    \email{o.s.schoyen@fys.uio.no}
    \affiliation{Department of Physics, University of Oslo, N-0316 Oslo, Norway}

    \author{H{\aa}kon Emil Kristiansen}
    \affiliation{Hylleraas Centre for Quantum Molecular Sciences, Department of
    Chemistry, University of Oslo, P.O. Box 1033 Blindern, N-0315 Oslo, Norway}

    \author{Alfred Alocias Mariadason}
    \affiliation{Expert Analytics AS, Tordenskiolds gate 6, 0160, Oslo, Norway}

    \date{\today}

    \begin{abstract}
        We have developed a McMurchie-Davidson-like recursion formula for
        efficient evaluation of the Coulomb attraction and interaction matrix
        elements between two-dimensional primitive Cartesian Gaussian type
        orbitals.
        We also present recurrence schemes for combined position and
        differential operator integrals, and three-center Gaussian integrals.
        The Cartesian Gaussian orbitals are isotropic in the exponent, but with
        arbitrary centers and angular momentum.
    \end{abstract}

    \maketitle

    \section{Introduction}
        Two-dimensional many-particle quantum mechanical systems are routinely
        studied both theoretically and experimentally.
        Systems such as lateral quantum dots and rings, \cite{divincenzo-3,
        xuedong, nielsen, reimann, chwiej, hjorth-jensen, popsueva} and
        electrons on liquid helium \cite{liquid-helium-platzman} are confined to
        planar motion and can be treated as two-dimensional structures.
        They are seen as promising candidate models for qubits, and can be used
        to build quantum computers.

        Solving these systems theoretically requires the use of approximations
        for all but the simplest problems.
        In particular for the case of multiparticle problems and arbitrary
        potential geometries.
        A large number of \emph{ab initio} many-body methods are formulated in
        terms of a known single-particle basis.
        For some problems the eigenstates of the single-particle Hamiltonian
        can be used, but these states can make the solution of the interaction
        terms more involved.
        In three dimensions the usage of Cartesian Gaussian type orbitals have
        been the prevalent choice for \emph{ab initio} methods applied to atomic
        and molecular systems \cite{helgaker2014molecular}.
        The main reason for this is that there exists several efficient
        algorithms \cite{mcmurchie, obara} for the evaluation of matrix
        elements between Gaussian type functions and relevant one- and two-body
        operators.
        In particular, the computationally expensive, non-separable Coulomb
        attraction and interaction integrals can be expressed in terms of fast
        recursion relations for arbitrary angular momentum values.

        However, due to the non-separability of the Coulomb integrals, the
        results for three dimensions do not generalize to lower dimensions.
        In particular, the computationally most intensive task of computing the
        two-body Coulomb interaction integrals are completely different in the
        two-dimensional case compared to the three-dimensional one.
        A common basis in two dimensions are the polar harmonic oscillator
        eigenstates, or the Fock-Darwin states, as the Coulomb interaction
        integrals can be solved analytically \cite{girvin-jach, anisimovas}.
        These are well suited for a single potential well located at the origin.
        In more complicated geometries with several well-centers, e.g., a double
        quantum dot, the quality of this basis rapidly deteriorates.

        We propose the use of primitive Cartesian Gaussian type orbitals in
        two-dimensions as they are much more flexible in their shape and
        placement allowing for a wider range of potential geometries.
        In two dimensions the Coulomb integrals between $s$-Gaussians, i.e.,
        spherical charge distributions, have a closed-form solution in terms of
        a modified Bessel function \cite{nielsen}.
        Previous work has often utilized several $s$-Gaussian orbitals to
        create a good enough single-particle basis \cite{xuedong, nielsen}.
        Using $p$, $d$, and higher-order Gaussians gives a better
        single-particle basis and lowers the number of states needed to build a
        sufficiently good representation of the problem.
        In this article, we have extended these solutions by introducing
        McMurchie-Davidson-like recursion relations for the evaluation of
        Coulomb integrals between Gaussian type orbitals of arbitrary angular
        momentum.
        This creates a general framework for studying two-dimensional systems
        using \emph{ab initio} methods.

        In \autoref{sec:theory} we define the primitive Cartesian Gaussian type
        orbitals and their properties.
        We also demonstrate how they can be expanded in Hermite-Gaussian
        functions and list the overlap integral between two Gaussian orbitals.
        We proceed to present recursion relations for the calculation of matrix
        elements between two Gaussian orbitals and the combined position and
        differential operator \footnote{
            We use the differential operator instead of the momentum operator
            to ensure that the matrix elements are strictly real-valued.
        }, and a ``Gaussian well''-operator \footnote{
            This is also known as a three-center integral.
        }.
        Next, we derive the recursion relations for the integral over two
        Gaussian orbitals and the Coulomb attraction operator, before we
        demonstrate how these relations can be reused for the integral over
        four Gaussian orbitals and the Coulomb interaction operator.
        Finally, we conclude with some remarks on the application of these basis
        functions and future prospects.

    \section{Theory}
        \label{sec:theory}
        A two-dimensional primitive unormalized Cartesian Gaussian orbital may
        be written \cite{helgaker2014molecular}
        \begin{align}
            g_{\alpha}(\vb{r}; a, \vb{A})
            = x^{i}_{A} y^{k}_{A} \exp(-a r^2_{A}),
        \end{align}
        where $\alpha = (i, k) \in \mathbb{N}^2_0$ are the
        Cartesian quantum numbers with the total orbital angular moment given
        by $l = i + k$, $\vb{r}_A \equiv \vb{r} - \vb{A}$, $r_A =
        \norm{\vb{r}_A}$ with $\vb{A} = (A_x, A_y) \in \mathbb{R}^{2}$ the
        center of the Gaussian, and $a \in \mathbb{R} > 0$ as the spread of the
        Gaussian.
        We will restrict $a$ be a scalar making the Gaussians separable in each
        Cartesian direction, viz.  $g_{\alpha}(\vb{r}; a, \vb{A}) = g_{i}(x; a,
        A_x) g_{j}(y; a, A_y)$.
        This lets us calculate many of the integrals as products of
        one-dimensional Gaussians.
        Multiplying two one-dimensional Gaussians results in a new Gaussian
        called the \emph{overlap distribution},
        \begin{align}
            g_{i}(x; a, A) g_{j}(x; b, B)
            &= K x^{i}_{A} x^{j}_{B}
            \exp(-p x^2_{P})
            \nonumber \\
            &\equiv
            \Omega_{ij}(x; p, \mu, P, X_{AB}),
        \end{align}
        where the parameters are the total exponent $p = a + b$, the center of
        mass $P = (aA + bB) / p$, $K \equiv \exp(-\mu X^2_{AB})$ with the
        reduced exponent $\mu = ab / (a + b)$, and the center difference
        $X_{AB} = A - B$.
        For notational brevity, we will not include the parameters, i.e., $p$,
        $\mu$, $P$, and $X_{AB}$, when writing the overlap distribution.
        It should be understood that the parameters for $\Omega_{ij}(x)$ are
        built from $g_i(x)$ and $g_j(x)$.
        In two dimensions the overlap distribution is given by a product of the
        overlap distribution between the $x$- and $y$-components of the
        Gaussian functions, viz., $g_{\alpha}(\vb{r})g_{\beta}(\vb{r}) =
        \Omega_{\alpha\beta}(\vb{r}) = \Omega_{ij}(x)\Omega_{kl}(y)$, where
        $\alpha = (i, k)$ and $\beta = (j, l)$, and the $x$- and $y$-components
        of the center of mass $\vb{P} = (P_x, P_y)$ and center difference
        $\vb{R}_{AB} = \vb{A} - \vb{B} = (X_{AB}, Y_{AB})$ are parameters to
        $\Omega_{ij}(x)$ and $\Omega_{kl}(y)$ separately.

        Now, as in the McMurchie-Davidson scheme \cite{mcmurchie,
        helgaker2014molecular} we expand the overlap distributions in a linear
        combination of Hermite-Gaussian functions,
        \begin{align}
            \Omega_{ij}(x)
            = \sum_{t = 0}^{i + j} E^{ij}_{t} \Lambda_{t}(x),
        \end{align}
        where $E^{ij}_{t}(p, \mu, P, X_{AB})$ are the expansion coefficients
        and $\Lambda_{t}(x; p, P)$ the Hermite-Gaussian functions given by
        \begin{align}
            \Lambda_{t}(x; p, P)
            = \pdv[t]{}{P} \exp(-p x_P^2),
        \end{align}
        where the parameters $p$, $\mu$, $P$ and $X_{AB}$ in the expansion
        coefficients and the Hermite-Gaussian functions are the parameters from
        the overlap distribution.
        This expansion is exact as the overlap distribution $\Omega_{ij}(x)$ is
        a polynomial of degree $i + j$ in $x_P$ and we expand it in a linear
        combination of Hermite-Gaussian functions of degree $t \leq i + j$ in
        $x_P$ \cite{helgaker2014molecular}.
        The two-dimensional expansion is then a product of the two
        one-dimensional expansions
        \begin{align}
            \Omega_{\alpha\beta}(\vb{r})
            &=
            \sum_{t = 0}^{i + j} \sum_{u = 0}^{k + l}
            E^{\alpha \beta}_{tu}
            \Lambda_{tu}(\vb{r}),
            \label{eq:od-hg-expansion}
        \end{align}
        with $E^{\alpha\beta}_{tu} = E^{ij}_{t}E^{kl}_{u}$ and
        $\Lambda_{tu}(\vb{r}) = \Lambda_{t}(x)\Lambda_{u}(y)$.
        The recurrence scheme for the expansion coefficients are given by
        \begin{gather}
            E^{i + 1, j}_{t}
            = \frac{1}{2p} E^{ij}_{t - 1}
            + X_{PA} E^{ij}_{t}
            + (t + 1) E^{ij}_{t + 1}, \\
            E^{i, j + 1}_{t}
            = \frac{1}{2p} E^{ij}_{t - 1}
            + X_{PB} E^{ij}_{t}
            + (t + 1) E^{ij}_{t + 1},
        \end{gather}
        where $E^{00}_{0} = K= \exp(-\mu X^{2}_{AB})$, and $E^{ij}_{t} = 0$ for
        $t < 0$ and $t > i + j$, with the derivation being listed in appendix
        \ref{app:od-hg-expansion}.
        The overlap integral between two one-dimensional Gaussian primitives
        can now be found by using the expansion of the overlap distribution as
        Hermite-Gaussian functions,
        \begin{align}
            s_{ij}
            &= \braket{g_i(a, A)}{g_j(b, B)}
            = E^{ij}_{0} \sqrt{\frac{\pi}{p}},
            \label{eq:overlap-integral}
        \end{align}
        where $p = a + b$ is the total exponent.
        The derivation of this expression can be found in appendix
        \ref{app:overlap-integrals}.
        The normalization factor of a one-dimensional primitive Gaussian
        orbital $g_i(x; a, A)$ is found by $N = \sqrt{s_{ii}}$.
        Extending to two dimensions is done by evaluating a product of two
        overlap integrals; one for each Cartesian direction, viz.,
        \begin{align}
            s_{\alpha\beta}
            = \braket{g_{\alpha}}{g_{\beta}}
            = \braket{g_{ik}}{g_{jl}}
            = s_{ij} s_{jl}.
        \end{align}

        \subsection{Combined position and differential operator integral}
            The second integral we wish to solve using arbitrary primitive
            Gaussian functions is the combined position and differential
            operator.
            We derive this expression for the differential operator instead of
            the momentum operator to ensure that all integrals are real.
            To ensure full generality we allow for arbitrary centers of the
            position operator, and arbitrary exponents for both the position
            and differential operator.
            The integral we wish to solve is thus
            \begin{align}
                L^{ef}_{ij}
                &= \mel{g_i(a, A)}{\hat{x}^{e}_C \dv[f]{}{x}}{g_j(b, B)},
                \label{eq:L-ef-ij}
            \end{align}
            with extensions to two dimensions by taking the product of a
            similar integral with $x \to y$.
            There are two downward recursion formulas – one for the position
            term and another for the differential operator – for this
            integral.
            They are given by
            \begin{gather}
                L^{e + 1, f}_{ij}
                = L^{ef}_{i + 1, j}
                + X_{AC} L^{ef}_{ij},
                \label{eq:x-mm-do-integral}
                \\
                L^{e, f + 1}_{ij}
                = j L^{ef}_{i, j - 1}
                - 2b L^{ef}_{i, j + 1},
                \label{eq:diff-mm-do-integral}
            \end{gather}
            where $X_{AC} \equiv A - C$.
            The base case of the recursion is the overlap integral, i.e.,
            $L^{00}_{ij} = s_{ij}$ from \autoref{eq:overlap-integral}.
            See appendix \ref{app:mm-do-integrals} for the derivation – which
            is based on the work of \citet{tellgren} – of these matrix
            elements.
            As an example, from \autoref{eq:L-ef-ij} we can build the kinetic
            operator matrix elements in two dimensions by
            \begin{align}
                T_{\alpha\beta}
                &= \mel{g_{\alpha}}{
                    \left(\dv[2]{}{x} + \dv[2]{}{y}\right)
                }{g_{\beta}}
                \nonumber \\
                &= L^{02}_{ij}s_{kl}
                + s_{ij}L^{02}_{kl},
            \end{align}
            where $\alpha = (i, k)$, $\beta = (j, l)$, and we have inserted the
            one-dimensional overlap integral between the Cartesian components
            that are not dependent on the derivative operator.

        \subsection{Gaussian well integral}
            When modeling $n$-well systems it is useful to tune the placement
            and shape of the various wells.
            One of the easiest potential wells that allow for this flexibility
            are Gaussian potential wells.
            This results in three-center integrals \cite{obara}.
            For the sake of full generality we therefore define the
            one-dimensional Gaussian potential well to be on the form
            \begin{align}
                \hat{v}_k(x; c, C, w)
                = -w x_C^{k} \exp(-c x_C^2),
            \end{align}
            that is, the Gaussian well is itself a primitive Gaussian with an
            exponent $c > 0$ and $w > 0$ giving a negative weight.
            Extending to higher dimensions results in a product between two
            one-dimensional wells.
            The matrix elements for this operator is given by
            \begin{align}
                G^{k}_{ij}
                &= \mel{g_i(a, A)}{\hat{v}_k}{g_j(b, B)}
                = -w\sum_{t = 0}^{i + j}
                E^{ij}_{t} P^{t}_{k0},
            \end{align}
            where $E^{ij}_{t}$ are the Hermite-Gaussian expansion coefficients
            between $g_i(x; a, A)$ and $g_j(x; b, B)$, and $P^{t}_{kl}$ can be
            found from the recurrence scheme
            \begin{align}
                P^{t}_{kl} &= \mel{g_k(c, C)}{\pdv[t]{}{P}}{g_l(p, P)}
                \nonumber \\
                &= 2p P^{t - 1}_{k, l + 1} - l P^{t - 1}_{k, l - 1},
            \end{align}
            with the base case $P^{0}_{kl} = \braket{g_k(c, C)}{g_l(p, P)} =
            s_{kl}$, i.e., the overlap integral.
            It is worth noting which Gaussians are included in the recurrence
            scheme for $P^t_{kl}$ as this is a two-center integral.
            The $g_l(x; p, P)$ Gaussian is the result from the product of
            $g_i(x; a, A)$ and $g_j(x; b, B)$ with $p$ as the total exponent
            and $P$ the center of mass from the product.
            The Gaussian $g_k(x; c, C)$ is the potential well.
            The derivation of this expression can be found in appendix
            \ref{app:derivation-gaussian-well-scheme}.

        \subsection{Coulomb attraction integral}
            Next, we look at the Coulomb attraction integral in two dimensions.
            This integral is not separable in the two Cartesian directions and
            we have to treat the integral in full.
            We wish to compute
            \begin{align}
                v_{\alpha \beta}
                &= \mel{g_{\alpha}(a, \vb{A})}{
                    \frac{1}{r_C}
                }{g_{\beta}(b, \vb{B})}
                = \int\dd[2]{\vb{r}}
                \frac{
                    \Omega_{\alpha\beta}(\vb{r})
                }{
                    r_C
                },
            \end{align}
            where $r_C \equiv \norm{\vb{r} - \vb{C}}$, and $\alpha = (i, k)$
            and $\beta = (j, l)$ are the compound indices of the
            two-dimensional Gaussians.
            We have set $e = (4\pi \epsilon_0)^{-1} = 1$ to keep the
            expressions as clean as possible.
            We expand the Gaussian overlap distribution in Hermite-Gaussian
            functions and move the derivative operators in front of the
            integral.
            This gives
            \begin{align}
                v_{\alpha \beta}
                &=
                \sum_{tu} E^{\alpha \beta}_{tu}
                \pdv[t]{}{P_x} \pdv[u]{}{P_y}
                \int \dd[2]{\vb{r}}
                \frac{\exp(-p r^{2}_{P})}{r_C},
                \label{eq:coulomb-attraction-integral-1}
            \end{align}
            where $\sum_{tu} \equiv \sum_{t = 0}^{i + j} \sum_{u = 0}^{k + l}$.
            The closed form expression for the two-dimensional integral over the
            spherical Gaussian is derived in appendix
            \ref{app:coulomb-attraction-s-gauss}, and is given by
            \begin{align}
                \int \dd[2]{\vb{r}}
                \frac{\exp(-p r^{2}_{P})}{r_C}
                = \pi\sqrt{\frac{\pi}{p}}
                \extbes_0\!\left(-p \Sigma^2/2\right),
                \label{eq:coulomb-attraction-s-gauss}
            \end{align}
            where $\extbes_n(z) \equiv \exp(z) I_n(z)$ is the
            \emph{exponentially scaled, modified Bessel function of the first
            kind}, and $\vb{\Sigma} = \vb{P} - \vb{C}$.
            Inserted back into \autoref{eq:coulomb-attraction-integral-1} we
            have
            \begin{align}
                v_{\alpha \beta}
                &=
                \pi \sqrt{\frac{\pi}{p}}
                \sum_{tu} E^{\alpha \beta}_{tu}
                \tilde{I}_{tu}\!\left(
                    -p\Sigma^2/2
                \right),
                \label{eq:coulomb-attraction-int}
            \end{align}
            where we have defined
            \begin{align}
                \tilde{I}_{tu}\!\left(
                    -p\Sigma^2/2
                \right)
                &\equiv
                \pdv[t]{}{P_x}
                \pdv[u]{}{P_y}
                \extbes_0\!\left(
                    -p\Sigma^2/2
                \right),
                \label{eq:deriv-ext-bes}
            \end{align}
            Now the task is to find a recurrence formula for the evaluation of
            \autoref{eq:deriv-ext-bes}.
            We therefore introduce the intermediate integrals
            \begin{gather}
                \tilde{I}^{n}_{tu}\!\left(
                    -p\Sigma^2/2
                \right)
                \equiv
                \pdv[t]{}{P_x}
                \pdv[u]{}{P_y}
                \extbes_n\!\left(-p\Sigma^2/2\right),
                \label{eq:intermediate-bessel-rec}
            \end{gather}
            where $\tilde{I}_{tu}(z) \equiv \tilde{I}^{0}_{tu}(z)$ and
            $\tilde{I}^{n}_{00}(z) \equiv \extbes_{n}(z)$.
            Incrementing $t$ and $u$ (separately) by one in
            \autoref{eq:intermediate-bessel-rec}, and collecting terms we find
            the relations
            \begin{align}
                \tilde{I}^{n}_{t + 1, u}
                &=
                -\frac{p}{2}\Biggl[
                    t\left(
                        \tilde{I}^{n - 1}_{t - 1, u}
                        + 2 \tilde{I}^{n}_{t - 1, u}
                        + \tilde{I}^{n + 1}_{t - 1, u}
                    \right)
                    \nonumber \\
                    &\qquad
                    +
                    \Sigma_x\left(
                        \tilde{I}^{n - 1}_{tu}
                        + 2\tilde{I}^{n}_{tu}
                        + \tilde{I}^{n + 1}_{tu}
                    \right)
                \Biggr],
                \label{eq:I-tilde-rec-attr-x}
                \\
                \tilde{I}^{n}_{t, u + 1}
                &=
                -\frac{p}{2}\Biggl[
                    u\left(
                        \tilde{I}^{n - 1}_{t, u - 1}
                        + 2 \tilde{I}^{n}_{t, u - 1}
                        + \tilde{I}^{n + 1}_{t, u - 1}
                    \right)
                    \nonumber \\
                    &\qquad
                    +
                    \Sigma_y\left(
                        \tilde{I}^{n - 1}_{tu}
                        + 2\tilde{I}^{n}_{tu}
                        + \tilde{I}^{n + 1}_{tu}
                    \right)
                \Biggr],
                \label{eq:I-tilde-rec-attr-y}
            \end{align}
            where we have removed the function arguments (they are the same as
            in \autoref{eq:coulomb-attraction-int}) for brevity.
            These expressions apply for $n \geq 0$ and with $\extbes_{-1}(z) =
            \extbes_{1}(z)$.
            The recursion schemas are derived in
            \ref{app:coulomb-attraction-recursion}.

            The Coulomb attraction potential is seldom used in two-dimensional
            systems.
            They are similar to the Gaussian wells in that they are flexible in
            their placement and strength.
            We have included it in this article as recursion relations can be
            completely reused in the two-body interaction integral.

        \subsection{Coulomb interaction integral}
            We are now in a position to set up the Coulomb interaction matrix
            elements between two two-dimensional primitive Cartesian Gaussian
            functions.
            The Coulomb interaction operator is given by
            \begin{align}
                \hat{u}(\vb{r}_1, \vb{r}_2)
                = \frac{1}{\norm{\vb{r}_1 - \vb{r}_2}},
                \label{eq:coulomb-operator}
            \end{align}
            where we again have set all physical constants to unity, and we
            define $r_{12} \equiv \norm{\vb{r}_1 - \vb{r}_2}$.
            The matrix elements are then given by
            \begin{align}
                u^{\alpha\beta}_{\gamma\delta}
                &\equiv
                \mel{g_{\alpha}(a, \vb{A}) g_{\beta}(b, \vb{B})}{
                    \hat{u}
                }{g_{\gamma}(c, \vb{C}) g_{\delta}(d, \vb{D})}
                \nonumber \\
                &=
                \int \dd[2]{\vb{r}_1} \dd[2]{\vb{r}_2}
                \frac{
                    \Omega_{\alpha\gamma}(
                        \vb{r}_1
                    )
                    \Omega_{\beta\delta}(
                        \vb{r}_2
                    )
                }{
                    r_{12}
                }
                \nonumber \\
                &=
                \sum_{tu, \tau \nu} E^{\alpha\gamma}_{tu}
                E^{\beta\delta}_{\tau\nu}
                \int \dd[2]{\vb{r}_1} \dd[2]{\vb{r}_2}
                \frac{
                    \Lambda_{tu}(
                        \vb{r}_1
                    )
                    \Lambda_{\tau\nu}(
                        \vb{r}_2
                    )
                }{
                    r_{12}
                }.
            \end{align}
            For the integrals over the Hermite-Gaussians (sans the expansion
            coefficients) we have
            \begin{align}
                V_{tu; \tau \nu}
                &=
                \pdv[t]{}{P_x}\pdv[u]{}{P_y}
                \pdv[\tau]{}{Q_x}\pdv[\nu]{}{Q_y}
                V_0,
            \end{align}
            where we have denoted the integral over the two spherical Gaussian
            charge distributions by $V_0$.
            This integral has a closed form solution given by
            \begin{align}
                V_0
                &=
                \int \dd[2]{\vb{r}_1} \dd[2]{\vb{r}_2}
                \frac{
                    \exp(-p\vb{r}^2_{1P})
                    \exp(-q\vb{r}^2_{2Q})
                }{
                    r_{12}
                }
                \nonumber \\
                &=
                \frac{\pi^2}{pq}\sqrt{\frac{\pi}{4 \sigma}}
                \extbes_{0}\!\left(-\Delta^2/(8\sigma)\right),
            \end{align}
            with $\vb{r}_{1P} \equiv \vb{r}_1 - \vb{P}$ and similarly for
            $\vb{r}_{2Q}$, $\sigma = (p + q) / (4pq)$, and $\vb{\Delta} =
            \vb{Q} - \vb{P}$ \cite{nielsen}.
            This derivation is demonstrated in appendix
            \ref{app:coulomb-s-gauss}.
            Except for the function argument and the prefactor, the Coulomb
            interaction integral between two spherical Gaussian orbitals is
            identical to the Coulomb attraction integral.
            To be able to reuse the recursion formula in
            \autoref{eq:I-tilde-rec-attr-x} we look at the function argument to
            the exponentially scaled Bessel function.
            Defining $f \equiv -\Delta^2/(8 \sigma)$ for the function argument,
            we have that
            \begin{align}
                \pdv[]{f}{Q_i}
                = -\frac{\Delta_i}{4\sigma}
                = -\pdv[]{f}{P_i},
                \quad
                \pdv[2]{f}{Q_i}
                = \pdv[2]{f}{P_i}
                = -\frac{1}{4\sigma}.
            \end{align}
            This lets us write the integral over the two Hermite-Gaussian
            functions as
            \begin{align}
                V_{tu; \tau \nu}
                &=
                (-1)^{t + u}
                \frac{\pi^2}{pq}\sqrt{\frac{\pi}{4\sigma}}
                \tilde{I}_{t + \tau, u + \nu}\!\left(
                    -\Delta^2/(8\sigma)
                \right),
            \end{align}
            where $\tilde{I}_{t + \tau, u + \nu}$ is the same as in
            \autoref{eq:deriv-ext-bes} with $P_i \to Q_i$ in the center
            derivatives shown in \autoref{eq:intermediate-bessel-rec} \footnote{
                This is because $\vb{\Delta} = \vb{Q} - \vb{P}$, i.e., $\vb{Q}$
                is the positive term in the center difference for the Coulomb
                interaction.
                For the Coulomb attraction integral this definition is reversed
                with $\vb{\Sigma} = \vb{P} - \vb{C}$.
            }.
            We therefore find the same recursion formula as in
            \autoref{eq:I-tilde-rec-attr-x}, but with $p \to (4\sigma)^{-1}$ and
            $\Sigma_i \to \Delta_i$.
            This leaves us with the full expression for the Coulomb interaction
            matrix elements
            \begin{align}
                u^{\alpha \beta}_{\gamma \delta}
                &=
                \frac{\pi^2}{pq}\sqrt{\frac{\pi}{4\sigma}}
                \sum_{tu}(-1)^{t + u} E^{\alpha \gamma}_{tu}
                \nonumber \\
                &\quad \times
                \sum_{\tau \nu} E^{\beta \delta}_{\tau \nu}
                \tilde{I}_{t + \tau, u + \nu}\!\left(
                    -\Delta^2/(8 \sigma)
                \right).
            \end{align}

    \section{Summary remarks}
        In this report, we have derived recursion formulas for the evaluation of
        atomic integrals using two-dimensional primitive Gaussian orbitals.
        The formulas are an extension of the three-dimensional
        McMurchie-Davidson method to two-dimensional systems.
        These integrals allow for more flexibility in the construction of basis
        sets for \emph{ab initio} methods applied to two-dimensional systems,
        e.g., lateral quantum dots.

        In this paper, we have only been concerned with the technicalities of
        computing matrix elements, but studies as to how the Gaussian basis
        elements should be chosen, e.g., placement of the centers, width of the
        peak, is of interest.
        Much can be learned from previous work on contracted primitive Gaussian
        orbitals in the three-dimensional case.

    \begin{acknowledgments}
        This work was supported by the Research Council of Norway (RCN) through
        its Centres of Excellence scheme,	project number 262695.
    \end{acknowledgments}

    \appendix

    \section{Hermite-Gaussian expansion coefficients}
        \label{app:od-hg-expansion}
        The expansion coefficients $E^{\alpha\beta}_{tu}$ in
        \autoref{eq:od-hg-expansion} are separable in each Cartesian direction.
        That is,
        \begin{align}
            E^{\alpha \beta}_{tu}
            = E^{ij}_{t}
            E^{kl}_{u},
        \end{align}
        with $\alpha = (i, k)$ and $\beta = (j, l)$.
        Restricting our attention to the one-dimensional case we expand the
        overlap distribution in Hermite-Gaussians
        \begin{align}
            \Omega_{ij}(x)
            = g_i(x; a, A) g_j(x; b, B)
            = \sum_{t = 0}^{i + j}
            E^{ij}_{t} \Lambda_{t}(x),
        \end{align}
        where this will be an exact expansion as $\Omega_{ij}(x)$ is a
        polyomial of degree $i + j$ in $x_P$.
        To build a recurrence scheme we increment one of the Gaussians in the
        overlap distribution and write
        \begin{align}
            \Omega_{i + 1, j}(x)
            &=
            g_{i + 1}(x) g_j(x)
            =
            x_A g_{i}(x) g_j(x)
            \nonumber \\
            &=
            x_P \Omega_{ij}(x) + X_{PA} \Omega_{ij}(x).
            \label{eq:overlap-dist-rec-1}
        \end{align}
        Looking at the first term, we expand the overlap distribution in
        Hermite-Gaussian functions.
        We can then find a relation for the Hermite-Gaussian multiplied by
        $x_P$, viz.,
        \begin{align}
            x_P \Lambda_t(x)
            &=
            x_P \pdv[t]{}{P} \exp(-p x_P^2)
            \nonumber \\
            &= \frac{1}{2p} \Lambda_{t + 1}(x) + t \Lambda_{t - 1}(x),
        \end{align}
        where we have used the commutation relation
        \begin{align}
            \left[
                x_P, \pdv[t]{}{P}
            \right]
            = t \pdv[t - 1]{}{P}.
        \end{align}
        Inserted back into \autoref{eq:overlap-dist-rec-1} (and removing
        function arguments) we get
        \begin{align}
            \Omega_{i + 1, j}
            &= \sum_{t = 0}^{i + j + 1} E^{i + 1, j}_{t} \Lambda_{t}
            = \left[x_P + X_{PA}\right] \Omega_{ij}
            \nonumber \\
            &= \sum_{t = 0}^{i + j} E^{ij}_{t} \left[
                \frac{1}{2p}\Lambda_{t + 1}
                + X_{PA} \Lambda_{t}
                + t \Lambda_{t - 1}
            \right]
            \nonumber \\
            &=
            \sum_{t = 1}^{i + j + 1} \frac{1}{2p} E^{ij}_{t - 1} \Lambda_{t}
            + \sum_{t = -1}^{i + j - 1} (t + 1) E^{ij}_{t + 1} \Lambda_{t}
            \nonumber \\
            &\qquad
            + \sum_{t = 0}^{i + j} X_{PA} E^{ij}_{t} \Lambda_{t},
        \end{align}
        where we have changed the limits on the two first sums.
        Now we wish to collect all terms under a common sum and factor out the
        Hermite-Gaussian functions on both sides of the equation.
        First, we require that $E^{ij}_{t} = 0$ for $t < 0$ allowing the
        starting limit on the first sum to be set to $0$.
        We can also set $t = 0$ to the starting limit for the second sum as the
        $(t + 1)$-term removes the $t = -1$ limit.
        Second, as we restrict our attention to polynomials of at most degree
        $i + j$ \footnote{
            Recall that we are working with an expansion of $\Omega_{ij}$ on
            the right-hand side.
            The left-hand side describes an expansion of polynomials of degree
            $i + j + 1$.
        } we require that $E^{ij}_{t} = 0$ for $t > i + j$.
        This lets us shift the limits on the second and third sums to $i + j +
        1$.
        Collecting we then get
        \begin{align}
            \Omega_{i + 1, j}
            &= \sum_{t = 0}^{i + j + 1}
            E^{i + 1, j}_{t} \Lambda_{t}
            \nonumber \\
            &=
            \sum_{t = 0}^{i + j + 1} \left[
                \frac{1}{2p} E^{ij}_{t - 1}
                + (t + 1) E^{ij}_{t + 1}
                + X_{PA} E^{ij}_{t}
            \right] \Lambda_{t}.
        \end{align}
        Equating terms at each order of Hermite-Gaussian then gives the
        recurrence scheme for the expansion coefficients at every step in the
        sum.
        We are then left with
        \begin{gather}
            E^{i + 1, j}_{t}
            = \frac{1}{2p} E^{ij}_{t - 1}
            + X_{PA} E^{ij}_{t}
            + (t + 1) E^{ij}_{t + 1}, \\
            E^{i, j + 1}_{t}
            = \frac{1}{2p} E^{ij}_{t - 1}
            + X_{PB} E^{ij}_{t}
            + (t + 1) E^{ij}_{t + 1},
        \end{gather}
        where $E^{00}_{0} = \exp(-\mu X^{2}_{AB})$, and $E^{ij}_{t} = 0$ for $t
        < 0$ and $t > i + j$.

    \section{Overlap integral between two Gaussian orbitals}
        \label{app:overlap-integrals}
        Here we derive the overlap integral – as seen in
        \autoref{eq:overlap-integral} – between two one-dimensional
        primitive Gaussians.
        \begin{align}
            s_{ij}
            &= \braket{g_i(a, A)}{g_j(b, B)}
            = \int \dd{x} g_i(x) g_j(x)
            \nonumber \\
            &= \int \dd{x} \Omega_{ij}(x)
            = \sum_{t = 0}^{i + j} E^{ij}_{t}
            \int \dd{x} \Lambda_{t}(x)
            \nonumber \\
            &= \sum_{t = 0}^{i + j} E^{ij}_{t}
            \pdv[t]{}{P}
            \int \dd{x} \exp(-p x^2_P)
            \nonumber \\
            &= \sum_{t = 0}^{i + j} E^{ij}_{t}
            \pdv[t]{}{P}
            \sqrt{\frac{\pi}{p}}
            = \sum_{t = 0}^{i + j} E^{ij}_{t}
            \delta_{t0}
            \sqrt{\frac{\pi}{p}}
            \nonumber \\
            &= E^{ij}_0
            \sqrt{\frac{\pi}{p}},
        \end{align}
        where $p$ is the total exponent from the product of the two Gaussian functions.
        The overlap between two two-dimensional primitive Gaussians will be a
        product of the overlap between each Cartesian direction.
        That is,
        \begin{align}
            s_{\alpha \beta}
            &= \braket{g_{\alpha}}{g_{\beta}}
            = \braket{g_{ik}}{g_{jl}}
            = s_{ij} s_{kl}.
        \end{align}

    \section{Combined position and differential operator integrals}
        \label{app:mm-do-integrals}
        We derive the recursion formulas for the combined position and
        differential operator integrals shown in \autoref{eq:x-mm-do-integral}
        and \autoref{eq:diff-mm-do-integral}.
        We start by finding a relation for the position term on a single
        Gaussian orbital.
        That is,
        \begin{align}
            g_i(x; a, A)x_C
            &= x^{i}_{A}\exp(-a x^2_A)\left(
                x - C - A + A
            \right)
            \nonumber \\
            &= x^{i}_{A}\exp(-a x^2_A)\left(
                x_A + X_{AC}
            \right)
            \nonumber \\
            &= g_{i + 1}(x)
            + X_{AC} g_{i}(x).
        \end{align}
        The downward recurrence scheme for the position is then
        \begin{align}
            L^{ef}_{ij}
            &= \mel*{g_i}{\hat{x}^{e}_{C} \dv[f]{}{x}}{g_j}
            \nonumber \\
            &=
            \Bigl[
                \bra*{g_{i + 1}}
                + X_{AC} \bra*{g_{i}}
            \Bigr]
            \hat{x}^{e - 1}_{C} \dv[f]{}{x}\!\ket*{g_j}
            \nonumber \\
            &= L^{e - 1, f}_{i + 1, j}
            + X_{AC} L^{e - 1, f}_{ij},
        \end{align}
        which corresponds to \autoref{eq:x-mm-do-integral}.
        The action of a derivative operator on a primitive one-dimensional
        Gaussian orbital is
        \begin{align}
            \dv[]{}{x}g_j(x; b, B)
            &= \dv[]{}{x_B} \left[
                x^i_B \exp(-b x^2_B)
            \right]
            \nonumber \\
            &=
            j g_{j - 1}(x)
            - 2b g_{j + 1}(x).
        \end{align}
        We thus find \autoref{eq:diff-mm-do-integral} from
        \begin{align}
            L^{ef}_{ij}
            &= \mel*{g_i}{\hat{x}^{e}_{C} \dv[f]{}{x}}{g_j}
            \nonumber \\
            &=
            \bra*{g_i}\!\hat{x}^{e}_{C} \dv[f - 1]{}{x} \Bigl[
                j \ket*{g_{j - 1}}
                - 2b \ket*{g_{j + 1}}
            \Bigr]
            \nonumber \\
            &= j L^{e, f - 1}_{i, j - 1}
            - 2b L^{e, f - 1}_{i, j + 1}.
        \end{align}

    \section{Gaussian well integral}
        \label{app:derivation-gaussian-well-scheme}
        The matrix element between two primitive Gaussians over a Gaussian well
        is given by
        \begin{align}
            G^{k}_{ij}
            &= \mel{g_i(a, A)}{\hat{v}_{k}(c, C, w)}{g_j(b, B)}
            \nonumber \\
            &=
            -w \int \dd{x} x_{C}^{k}
            \exp(-cx_{C}^{2})
            \Omega_{ij}(x)
            \nonumber \\
            &=
            -w \sum_{t = 0}^{i + j}
            E^{ij}_{t}
            \int \dd{x} g_k(x) \Lambda_{t}(x)
            \nonumber \\
            &=
            -w \sum_{t = 0}^{i + j}
            E^{ij}_{t}
            \mel{g_k(c, C)}{\pdv[t]{}{P}}{g_0(p, P)}
            \nonumber \\
            &=
            -w \sum_{t = 0}^{i + j}
            E^{ij}_{t}
            P^{t}_{k0}.
        \end{align}
        Here it is worth noting that the choice of creating an overlap
        distribution between $g_i(x; a A)$ and $g_j(x; b, B)$ is somewhat
        arbitrary.
        This can very well be done between the well and any of the two.
        However, this requires some care when manipulating the indices for the
        expansion coefficients and the matrix elements $P^{t}_{kl}$.
        We observe that $P^{0}_{kl} = \braket{g_k}{g_l} = s_{kl}$.
        For $t > 0$ we get
        \begin{align}
            P^{t}_{kl}
            &=
            \mel{g_k(c, C)}{\pdv[t]{}{P}}{g_l(p, P)}
            \nonumber \\
            &= \bra*{g_k}\!\pdv[t - 1]{}{P}\Bigl[
                2p \ket*{g_{l + 1}}
                - l \ket*{g_{l - 1}}
            \Bigr]
            \nonumber \\
            &=
            2p P^{t - 1}_{k, l + 1}
            - l P^{t - 1}_{k, l - 1}.
        \end{align}

    \section{Coulomb attraction integral over a spherical Gaussian charge
    distribution}
        \label{app:coulomb-attraction-s-gauss}
        We derive a closed-form expression for the Coulomb attraction integral
        over a two-dimensional spherical Gaussian charge distribution.
        We have
        \begin{align}
            v_{0}(C)
            &= \int \dd[2]{\vb{r}}
            \frac{\exp(-pr^2_P)}{r_C}
            \nonumber \\
            &= \int \dd[2]{\vb{r}}
            \exp(-pr^2_P)
            \int \dd{t}
            \frac{1}{\sqrt{\pi}}
            \exp(-r^2_{C} t^2),
        \end{align}
        where we in the last equation have used that
        \begin{align}
            \int\dd{t}\exp(-r^2_{C} t^2)
            = \frac{\sqrt{\pi}}{r_C},
        \end{align}
        for $\Re(r_C) > 0$, in reverse.
        We now use the Gaussian product rule
        \begin{align}
            \exp(-p r^2_{P})
            \exp(-r^2_{C} t^2)
            &=
            \exp(-\nu \Sigma^2)
            \exp(-s r^2_{S}),
        \end{align}
        where we have introducd $s = p + t^2$, $\nu = pt^2 / s$, $S =
        (p\vb{P} + t^2\vb{C})/(p + t^2)$, and $\vb{\Sigma} = \vb{P} - \vb{C}$.
        Inserting this back into the integral we can now solve the integral over
        $\dd[2]{\vb{r}}$ for the last term in the Gaussian product.
        This leaves us with
        \begin{align}
            v_{0}(C)
            &=
            \frac{1}{\sqrt{\pi}}
            \int\dd{t}
            \exp(-\frac{pt^2}{p + t^2} \Sigma^2)
            \frac{\pi}{p + t^2}.
        \end{align}
        This integral is symmetric for $t \in (-\infty, \infty)$.
        We therefore shift the limits of integration to $t \in [0, \infty)$ and
        multiply with a factor of $2$.
        Then
        \begin{align}
            v_{0}(C)
            &=
            \frac{2\sqrt{\pi}}{p}
            \int_{0}^{\infty} \dd{t}
            \frac{p}{p + t^2}
            \exp(-\frac{pt^2}{p + t^2} \Sigma^2),
        \end{align}
        where we have moved a factor of $p$ into the numerator of the integral.
        Next, we perform a variable substitution introducing
        \begin{gather}
            u^2 = \frac{t^2}{p + t^2}
            \implies
            u = \frac{t}{\sqrt{p + t^2}},
        \end{gather}
        where $t \geq 0$ from the integration limits and $\Re(p) > 0$.
        The new integration limits are $\lim_{t \to \infty} u(t) = 1$ and $u(0)
        = 0$.
        We also have
        \begin{gather}
            \dv[]{u}{t}
            = \frac{\left(1 - u^2\right)^{3 / 2}}{\sqrt{p}}
            \implies
            \dd{t} = \frac{\sqrt{p}}{\left(1 - u^2\right)^{3 / 2}} \dd{u},
        \end{gather}
        for the variable substitution.
        Inserted back into $v_{0}(C)$ we are then left with
        \begin{align}
            v_{0}(C)
            &=
            \frac{2\sqrt{\pi}}{p}
            \int_{0}^{1}
            \dd{u}\frac{\sqrt{p}\left(1 - u^2\right)}{\left(1 - u^2\right)^{3 / 2}}
            \exp(-p u^2 \Sigma^2)
            \nonumber \\
            &=
            \pi\sqrt{\frac{\pi}{p}}
            \exp(-\frac{p \Sigma^{2}}{2})
            I_0\!\left(-\frac{p \Sigma^{2}}{2}\right),
        \end{align}
        where $I_0(z)$ is the modified Bessel function of the first kind.
        Note that $I_0(z) = I_0(-z)$ making the choice of sign in the argument
        arbitrary.
        However, when we create a recursion formula for higher-order Hermite
        Gaussians we benefit from using the same sign in the exponential
        function and the Bessel function.

    \section{Recursion formula for the Coulomb attraction integral}
        \label{app:coulomb-attraction-recursion}
        Defining $z \equiv -p \Sigma^2 / 2$ as the argument of the exponential
        function and the Bessel function in
        \autoref{eq:coulomb-attraction-s-gauss}, we find the derivative of this
        quantity with respect to $P_x$ and $P_y$ to be
        \begin{gather}
            \pdv[]{z}{P_i}
            = -p\Sigma_i,
            \quad
            \pdv[2]{z}{P_i}
            = -p.
        \end{gather}
        Furthermore, we define the \emph{exponentially scaled modified Bessel
        function of the first kind} to be
        \begin{align}
            \extbes_n(z) \equiv \exp(z)I_n(z),
            \label{eq:extbes-def}
        \end{align}
        where $n$ denotes the order of the modified Bessel function.
        The derivative of this function is
        \begin{align}
            \left.\dv[]{\extbes_n}{z}\right\rvert_{z}
            &= \extbes_n(z) + \frac{1}{2}\left[
                \extbes_{n - 1}(z) + \extbes_{n + 1}(z)
            \right],
        \end{align}
        where $n \geq 0$ and $\extbes_{-1}(z) = \extbes_{1}(z)$.
        Using the intermediate integrals defined in
        \autoref{eq:intermediate-bessel-rec} and in the text below, we compute
        \begin{align}
            \tilde{I}^{n}_{t + 1, u}(p, \vb{\Sigma})
            &=
            \pdv[t]{}{P_x}
            \pdv[u]{}{P_y}
            \left.\pdv[]{\extbes_{n}}{P_x}\right\rvert_{z = -p\Sigma^2/2}
            \nonumber \\
            &=
            \pdv[t]{}{P_x}
            \pdv[u]{}{P_y}
            \pdv[]{z}{P_x}
            \left.\pdv[]{\extbes_{n}}{z}\right\rvert_{z = -p\Sigma^2/2}
            \nonumber \\
            &=
            \pdv[t]{}{P_x}
            \pdv[u]{}{P_y}
            \left(-p\Sigma_x\right)
            \frac{1}{2}\Bigl[
                \extbes_{n - 1}
                \nonumber \\
                &\qquad
                + 2 \extbes_{n} + \extbes_{n + 1}
            \Bigr],
        \end{align}
        where we for brevity have removed the function arguments to the
        exponentially scaled modified Bessel functions.
        To proceed from here we note that the derivative with respect to $P_y$
        ``passes through'' the term in front of the Bessel functions.
        For the derivatives with respect to $P_x$ we have that
        \begin{align}
            \left[\pdv[t]{}{P_x}, \Sigma_x\right]
            &= t \pdv[t - 1]{}{P_x}.
        \end{align}
        We then have
        \begin{align}
            \tilde{I}^{n}_{t + 1, u}(p, \vb{\Sigma})
            &=
            -\frac{p}{2} \left[
                t \pdv[t - 1]{}{P_x}
                + \Sigma_x \pdv[t]{}{P_x}
            \right]
            \nonumber \\
            &\qquad \times
            \left[
                \tilde{I}^{n - 1}_{0u}
                + 2 \tilde{I}^{n}_{0u}
                + \tilde{I}^{n + 1}_{0u}
            \right]
            \nonumber \\
            &=
            -\frac{p}{2}
            \Biggl[
                t \left(
                    \tilde{I}^{n - 1}_{t - 1, u}
                    + 2 \tilde{I}^{n}_{t - 1, u}
                    + \tilde{I}^{n + 1}_{t - 1, u}
                \right)
                \nonumber \\
                &\qquad
                + \Sigma_x \left(
                    \tilde{I}^{n - 1}_{t u}
                    + 2 \tilde{I}^{n}_{t u}
                    + \tilde{I}^{n + 1}_{t u}
                \right)
            \Biggr].
        \end{align}
        For the $y$-direction we have a similar relation given by
        \begin{align}
            \tilde{I}^{n}_{t, u + 1}(p, \vb{\Sigma})
            &=
            -\frac{p}{2}
            \Biggl[
                u \left(
                    \tilde{I}^{n - 1}_{t, u - 1}
                    + 2 \tilde{I}^{n}_{t, u - 1}
                    + \tilde{I}^{n + 1}_{t, u - 1}
                \right)
                \nonumber \\
                &\qquad
                + \Sigma_y \left(
                    \tilde{I}^{n - 1}_{t u}
                    + 2 \tilde{I}^{n}_{t u}
                    + \tilde{I}^{n + 1}_{t u}
                \right)
            \Biggr].
        \end{align}

    \section{Coulomb interaction between two spherical Gaussian charge
    distributions}
        \label{app:coulomb-s-gauss}
        We derive the closed-form expression for the Coulomb interaction
        integral between two two-dimensional spherical Gaussian charge
        distributions.
        This is based on the work done by \citeauthor{nielsen} \cite{nielsen}.
        The integral we wish to solve is on the form
        \begin{align}
            V_{0}
            &=
            \int \dd[2]{\vb{r}_1} \dd[2]{\vb{r}_2}
            \frac{\exp(-p r_{1P}^{2}) \exp(-q r_{2Q}^{2})}{r_{12}},
        \end{align}
        where we have defined $\vb{r}_{1P} \equiv \vb{r}_{1} - \vb{P}$ and
        similarly $\vb{r}_{2Q}$.
        We also have $r_{12} \equiv \norm{\vb{r}_1 - \vb{r}_2}$.
        To proceed we use the momentum representation of the exponential
        functions and the Coulomb operator.
        They are given by
        \begin{equation}
            \begin{gathered}
                \exp(-p r^{2}_{1P})
                =
                \frac{\pi}{p}\int\frac{\dd[2]{\vb{k}}}{(2\pi)^2}
                \exp(-\frac{k^2}{4p})\exp(-i\vb{k}\cdot\vb{r}_{1P}),
                \\
                \frac{1}{r_{12}}
                = \int \frac{\dd[2]{\vb{k}}}{(2\pi)^2}
                \frac{2\pi}{k}\exp(-i\vb{k}\cdot\vb{r}_{12}).
            \end{gathered}
        \end{equation}
        Inserted into the full integral we have
        \begin{align}
            V_{0}
            &=
            \frac{\pi^2}{pq}
            \int
            \dd[2]{\vb{r}_1} \dd[2]{\vb{r}_2}
            \frac{\dd[2]{\vb{k}_1}}{(2\pi)^2}
            \frac{\dd[2]{\vb{k}_2}}{(2\pi)^2}
            \frac{\dd[2]{\vb{k}_3}}{(2\pi)^2}
            \nonumber \\
            &\qquad\times
            \frac{2\pi}{k_2}\exp(-i\vb{k}_2\cdot\vb{r}_{12})
            \nonumber \\
            &\qquad\times
            \exp(-\frac{k_1^2}{4p})\exp(-i\vb{k}_1\cdot\vb{r}_{1P})
            \nonumber \\
            &\qquad\times
            \exp(-\frac{k_3^2}{4q})\exp(-i\vb{k}_3\cdot\vb{r}_{2Q}).
        \end{align}
        We collect all the $\vb{r}$-terms in the complex exponentials, and
        integrate over $\vb{r}_1$ and $\vb{r}_2$ yielding delta functions for
        the momentum integrals.
        We can then set $\vb{k} \equiv -\vb{k}_1 = \vb{k}_2 = \vb{k}_3$, and
        this lets us write
        \begin{align}
            V_0
            &= \frac{\pi^2}{pq}
            \int\frac{\dd[2]{\vb{k}}}{(2\pi)^2}
            \frac{2\pi}{k}
            \exp(-\sigma k^2)
            \exp(i\vb{k}\cdot\vb{\Delta}),
        \end{align}
        where we have defined $\sigma = (p + q) / (4pq)$ and $\vb{\Delta} =
        \vb{Q} - \vb{P}$.
        We now transform to polar coordinates and choose the $x$-axis to be
        along the $\vb{\Delta}$ such that we can write the scalar product
        between $\vb{k}$ and $\vb{\Delta}$ as $\vb{k}\cdot\vb{\Delta} =
        k\Delta\cos(\phi)$, in the normal way.
        We let $k \in (-\infty, \infty)$ and $\phi \in [0, \pi)$ as this
        covers the same area as the conventional limits.
        The integral can now be written
        \begin{align}
            V_0
            &= \frac{\pi}{2pq}
            \int_{0}^{\pi}\dd{\phi}
            \int_{-\infty}^{\infty}\dd{k}
            \exp(-\sigma k^2)
            \exp(ik\Delta\cos(\phi)),
        \end{align}
        where the Coulomb momentum term has been cancelled by the Jacobian from
        changing to polar coordinates.
        Completing the square in the exponentials we get
        \begin{align}
            -\sigma k^2 + ik\Delta\cos(\phi)
            &= -\sigma\left(k - \frac{i\Delta \cos(\phi)}{2\sigma}\right)^2
            \nonumber \\
            &\quad
            - \frac{\Delta^2\cos[2](\phi)}{4\sigma},
        \end{align}
        where only the squared term is dependent on $k$.
        We can then solve the $k$-integral leaving us with
        \begin{align}
            V_0
            &= \frac{\pi}{pq}
            \sqrt{\frac{\pi}{4\sigma}}
            \int_{0}^{\pi}\dd{\phi}
            \exp(-\frac{\Delta^2\cos[2](\phi)}{4\sigma}).
        \end{align}
        Next we rewrite the argument in the exponential using a trigonometric
        identity for double angles.
        That is
        \begin{align}
            -\frac{\Delta^2\cos[2](\phi)}{4\sigma}
            = -\frac{\Delta^2}{8\sigma}\bigl(1 + \cos(2\phi)\bigr),
        \end{align}
        and we move the constant term outside the integral.
        Defining $\theta = 2\phi$ we have $\dd{\theta} / 2 = \dd{\phi}$ and a
        doubling of the integral limits.
        We are then left with
        \begin{align}
            V_0
            &= \frac{\pi}{pq}
            \sqrt{\frac{\pi}{4\sigma}}
            \exp(-\frac{\Delta^2}{8\sigma})
            \int_{0}^{2\pi}\frac{\dd{\phi}}{2}
            \exp(-\frac{\Delta^2}{8\sigma}\cos(\theta)).
        \end{align}
        The last term in the integral is symmetric over the integration
        interval.
        This then gives
        \begin{align}
            V_0
            &= \frac{\pi}{pq}
            \sqrt{\frac{\pi}{4\sigma}}
            \exp(-\frac{\Delta^2}{8\sigma})
            \pi I_0\!\left(-\frac{\Delta^2}{8\sigma}\right),
        \end{align}
        with $I_0(z)$ being the modified Bessel function of the first kind.
        Collecting terms we are then left with
        \begin{align}
            V_0
            &= \frac{\pi^2}{pq}\sqrt{\frac{\pi}{4\sigma}}\extbes_0\!\left(
                -\frac{\Delta^2}{8\sigma}
            \right),
        \end{align}
        where we have inserted the definition of the exponentially scaled Bessel
        function from \autoref{eq:extbes-def}.

    \bibliography{references}

\end{document}